\documentstyle[12pt,epsfig]{article}   

\title{Elimination of ambiguities in $\pi\pi$ phase shifts
    using crossing symmetry}

\author{ R. Kami\'nski$^a$, L. Le\'sniak$^a$ and B. Loiseau$^b$ \vspace{0.3cm}\\
\hsl$^a$ Department of Theoretical Physics,\ H. Niewodnicza\'nski
Institute\\
of Nuclear Physics, PL 31-342 Krak\'ow, Poland\\
$^b$ Laboratoire de Physique Nucl\'eaire et de Hautes 
\'Energies\footnote{Unit\'e de Recherche des Universit\'es
Paris 6 et Paris 7, associ\'ee au CNRS}, Groupe \\Th\'eorie,
Univ. P. \& M. Curie, 4 Pl. Jussieu, F-75252 Paris, France }

\newcommand{\uf}{``up-flat" }   
\newcommand{\df}{``down-flat" }   
\newcommand{\pipi}{$\pi\pi$ }   
\newcommand{\mpp}{$m_{\pi\pi}$ }

\newcommand{\hst}{\hspace{-0.34cm} }   
\newcommand{\hsl}{\hspace{-0.3cm} }   
\newcommand{\hsmu}{\hspace{-0.5cm} }   

\begin{document}

\maketitle   

\begin{abstract}   

     Roy's equations, which 
    incorporate crossing symmetry of the 
    $\pi\pi$ scattering amplitudes, are used to 
    resolve the present ambiguity between two  
    solutions for the scalar-isoscalar phase shifts below 1~GeV. 
    It is shown that the ``down-flat" solution satisfies well Roy's 
    equations and consequently crossing symmetry while the other
    solution called ``up-flat" does not and thus 
    should be   eliminated.

\end{abstract}


\section{ Introduction }
\label{sec:introduction}

A good knowledge of the pion-pion scattering is important in studies of 
different processes in nuclear and particle physics~\cite{panic99}. 
Construction of phenomenological $\pi\pi$ amplitudes requires not only 
experimental input but also use of theoretical constraints such as 
unitarity, analyticity and crossing symmetry. 
Direct study of the $\pi\pi$ collisions is beyond present experimental 
possibilities. 
Phenomenological phase shifts are obtained through partial wave 
analyses of final states in which pions are produced. These analyses 
are often model dependent and can sometimes lead to ambiguous results.

In 1997 a study of the $\pi^-p_\uparrow\to\pi^+\pi^-n$ reaction on a 
polarized target has been performed in the $m_{\pi\pi}$ effective 
mass between 600  and 1600 MeV giving four solutions for the 
$\pi\pi$ scalar-isoscalar phase shifts below 1 GeV~\cite{kam97}. 
Using the unitarity constraint two ``steep" solutions were rejected and the 
two remaining ones, called ``down-flat" and ``up-flat", passed 
this test~\cite{kam00}. 
Elimination of this remaining ambiguity is necessary for extension of studies 
done near the $\pi\pi$ threshold and above using chiral perturbation 
models based on QCD (see for example \cite{gasser83, Amoros}).

A steady increase of the $\pi\pi$ scalar-isoscalar 
phase shifts below 1 GeV can be interpreted as due to the presence of a 
broad $\sigma$-meson~\cite{kyoto01,kam94}.
Previous elimination of ``steep" solutions excludes a narrow $\sigma$ 
with a width comparable to that of the $\rho$-meson~\cite{kam00}.
In this channel, near the $K\bar K$ threshold and above it, other 
resonances exist which could be mixed~with~glueball~states~predicted~by~QCD.

In order to eliminate the above mentioned ``up-down" ambiguity one can 
check if the corresponding amplitudes satisfy crossing symmetry. 
Roy's equations~\cite{roy71} can serve as a tool to perform this check.
They also correlate different experimental phase shifts determined 
near the $\pi\pi$ threshold and at higher energy, in particular for 
the scalar-isoscalar, $\ell=0,\ I=0$,
 the scalar-isotensor, $\ell=0,\ I=2$,
and the vector-isovector, $\ell=1,\ I=1$,
$\pi\pi$ partial waves.
These equations have been already applied by Pennington and 
Protopopescu~\cite{penn73} to study different $\pi\pi$ phase shift 
solutions obtained in the early seventies. 
They were able to resolve a ``steep-flat" type ambiguity~\cite{kam00} present in 
the phenomenological $\pi\pi$ amplitudes considered at that time.
Later two high-statistics experiments were 
performed~\cite{grayer75,becker79} and, as previously said, the analyses 
of these experiments lead to two plausible solutions
\cite{kam97,kam00}.
Recently a comprehensive analysis of Roy's equations for the $\pi\pi$ 
interaction has appeared~\cite{anan}. 
There, a special emphasis is put on the $m_{\pi\pi}$ range from the 
$\pi\pi$ threshold to 0.8 GeV. In the present paper we pay a particular
attention to the range between 0.8 and 1 GeV. 
In this range the largest differences 
between the ``up-flat" and ``down-flat" solutions occur, reaching 
values up to 45$^{o}$ (see Fig.~4 in~\cite{kam00} and Fig.~2~here).

\section{ Roy's equations }
\label{sec:roysequation}

Assuming analyticity one can write twice subtracted fixed-t
dispersion relations~\cite{roy71} for the unitary $\pi\pi$ amplitudes,

\begin{equation}
    \begin{array}{rcl}
        T^{I}(s,t,u) & = & 
	\displaystyle \sum_{I^\prime=0}^{2}C_{st}^{II^\prime}
        \left[
        B^{I^\prime}(t)+(s-u)D^{I^\prime}(t)
        \right] \\
         & + & \displaystyle {{1}\over {\pi}}
         \displaystyle \int\limits_{4\mu^2}^{\infty}
         \displaystyle \frac{ds'}{s^{\prime 2}}
         \left(
         \displaystyle \frac{s^2}{s'-s}\
         {1}^{II^\prime}
	+\displaystyle \frac{u^2}{s'-u}C_{su}^{II^\prime}
         \right)
         \mbox{Im }T^{I^\prime}(s',t,u').
    \end{array}
    \label{eq:1}
\end{equation}
In this equation $s,\ t$ and $u$ are the usual Mandelstam variables 
satisfying $s+t+u=4\mu^2$, where $\mu$ is the pion mass ; 
$C_{st}^{II^\prime}$ and $C_{su}^{II^\prime}$ are the isospin matrix 
elements and ${1}^{II^\prime}$ is the unit matrix.
The subtraction functions $B^{I^\prime}(t)$ and $D^{I^\prime}(t)$
can be related to the isospin 0 and 2 $S$-wave scattering lengths 
$a_{0}^{0}$ and $a_{0}^{2}$, respectively. 

Projection of (\ref{eq:1}) onto partial waves 
leads to Roy's equations,
\begin{equation}
    \begin{array}{rcl}
    \mbox{Re } f_{\ell}^{I}(s) & = &
        a_{0}^{0}\delta_{I0} \delta_{\ell 0} + a_{0}^{2}\delta_{I2}
	\delta_{\ell 0}\\
     & + &  \displaystyle \frac{s-4{\mu}^2}{12{\mu}^2}(2a_{0}^{0}-5a_{0}^{2})\    
       (\delta_{I0}\delta_{\ell 0}+
        \frac{1}{6}\delta_{I1}\delta_{\ell 1} 
        -\ \frac{1}{2}\ \delta_{I2}\delta_{\ell 0})\\
     & + & 
	\displaystyle \sum\limits_{I'=0}^{2}
	\displaystyle \sum\limits_{\ell'=0}^{1}
     \hspace{0.25cm}-\hspace{-0.55cm}
	\displaystyle \int \limits_{4\mu^2}^{s_{max}}\ ds'
     K_{\ell \ell^\prime}^{I I^\prime}(s,s') \mbox{Im }f_{\ell'}^{I^\prime}
     (s') + d_{\ell}^{I}(s,s_{max}).
\end{array}
 \label{eq:2}
\end{equation}
Here the functions $ K_{\ell \ell^\prime}^{I I^\prime}(s,s')$ are the 
kernels and $d_{\ell}^{I}(s,s_{max})$ are the so-called driving terms.
Detailed expressions of these functions can be found for instance 
in~\cite{anan,basd72}.
The driving terms contain the low-energy, $s'\leq s_{max}$, contributions 
from the partial waves $\ell^\prime\geq 2$ and the high-energy, 
$s^\prime\geq s_{max}$, contributions from all the partial waves.
If one assumes Mandelstam analyticity, the range of validity of Roy's 
equations~(\ref{eq:2}) extends to $68\mu^2=$ (1.15 GeV)$^2$~\cite{roy71}.

The partial waves  amplitudes  $f_{\ell}^{I}(s)$ are  related to the $\pi \pi$ phase shifts 
$\delta_{\ell}^{I}$ and inelasticities~$\eta_{\ell}^{I}$:
\begin{equation}
    f_{\ell}^{I}(s)=\sqrt{\frac{s}{s-4\mu^2}}\frac{1}{2i}
    \left(\eta_{\ell}^{I}e^{2i\delta_{\ell}^{I}}-1\right).
    \label{eq:3}
\end{equation}
Each set of experimentally determined phase shifts and 
inelasticities can serve as input to calculate the real and imaginary 
parts  of the partial wave amplitudes.
After a suitable parameterization, the imaginary parts can be inserted 
into Roy's equations~(\ref{eq:2}) from which one obtains the 
real parts as  {\it output}  to be compared with the corresponding 
real part  {\it input}.
The quantitative agreement between  output and input will be used 
to verify how well a given set of phase shifts satisfies  crossing 
symmetry in a particular range of $m_{\pi\pi}$ where Roy's equations 
can be applied.

\section{ Input amplitudes }
\label{inputamplitude}

The ``down-flat" and ``up-flat" data~\cite{kam97} together with the 
results on the $K\bar K$ phase shifts~\cite{cohen80} were analysed by 
us using a separable potential model of three coupled 
scalar-isoscalar channels ($\pi\pi,\ K\bar K$ and an effective $4\pi$ 
system)~\cite{kamplb97,kam99}.
This model yields particularly good fits to the ``down-flat" solution
from 600 to 1600 MeV. The ``up-flat" data are reasonably well 
described only above 970 MeV. 
From 800 to  970 MeV  the model values are too low in 
comparison with the ``up-flat" data (see Fig.~1b of~\cite{kamplb97}).
In our calculations we need a faithful representation
of both ``up-flat" and ``down-flat" solutions and 
therefore below 970 MeV we use the following Pad\'e approximation for the 
$\delta_{0}^0$ phase shifts:
\begin{equation}
    \tan\delta_{0}^0(s) = \frac{\sum_{i=0}^{4}\alpha_{2i+1}k^{2i+1}}
     {\Pi_{i=1}^{3}(k^2/\alpha_{2i}-1)}
    \label{eq:4}
\end{equation}
where $k=\frac{1}{2}\sqrt{s-4\mu^2}$ is the  pion  momentum 
and 
$\alpha_{j}\,(j=1,\ldots,7,9)$ are constant parameters. 
This choice is dictated by the analytical properties of the Jost functions.
These constants will be obtained from the best fits
to data and to Roy's equations for the \uf and \df separately. 
In these fits we also use the near
threshold phase-shift differences, $\delta_0^0-\delta_1^1$, recently
extracted from the high statistics $K_{e4}$ decay experiment~\cite{pislak01}.
Using the power expansion of the $S$-wave amplitudes near the $\pi\pi$ 
threshold,
\begin{equation}
    \mbox{Re }f_{0}^{I}(s)= a_{0}^{I}+b_{0}^{I}\ k^2+\ldots
    \label{eq:5},
\end{equation}
we can relate the parameters $a_{0}^0$ and $b_{0}^0$
to the constants $\alpha_{j}$. One has:
 $a_{0}^0 = -\alpha_{1}\mu $ and 
$b_{0}^0 = -\alpha_{1}\mu \left ( 0.5\mu^{-2}+\alpha_{2}^{-1}+\alpha_{4}^{-1}+\alpha_{6}^{-1}
-\alpha_{1}^2\right) - \alpha_{3}\mu$.
The parameters $\alpha_{7}$ and  $\alpha_{9}$ are chosen to 
match the values of $\delta_{0}^0$ following from the 
three-channel fits at 969 and 970~MeV~\cite{kamplb97}.
Above 970~MeV we use the fit A 
to describe the ``down-flat" solution and the fit C for the ``up-flat" one.
These model amplitudes are used up to  the $s_{max}$ value even if it exceeds 
1600 MeV  (see  (\ref{eq:2})).

The isotensor wave is parameterized within
 the separable potential model 
of~\cite{kam94, kamplb97}. We use a rank-two potential:
\begin{equation}
    V_{\pi\pi}^{I=2}(p,p')=\sum_{i=1}^{2}\lambda_{i}g_{i}(p)
    g_{i}(p'),
    \label{eq:6}
\end{equation}
where 
\begin{equation}
    g_{i}(p)=\sqrt{\frac{4\pi}{\mu}}\ \frac{1}{p^2+\beta_{i}^2}
    \label{eq:7}
\end{equation}
are form factors with range parameters $\beta_{i}$, $p$ and $p'$ are the pion center of
mass momenta in the initial and final states, respectively.
The threshold parameters $a^2_0$ and $b^2_0$ can be related to the strength parameters 
$\Lambda_{i}=\lambda_{i}/(2\beta_{i}^3)$.
The $I=2$ $\pi \pi$ phase shifts 
from~\cite{hoogland77}, obtained with their method A,
serve as input in our fitting procedure.
Here we assume the isotensor wave to be elastic ($\eta_0^2 \equiv 1$) from the \pipi threshold to 
$s_{max}$.

For the $P$-wave, from threshold to 970 MeV, we use an extended
Schenk parameterization as defined in~\cite{anan}

\begin{equation}
\tan \,\delta_1^1(s) = 
\sqrt{1-\frac{4\mu^2}{s}}
\ k^2\left(A+Bk^2+Ck^4+Dk^6\right)\left(\frac{4\mu^2-s_\rho}{s-s_\rho}\right).
\label{eq:schenk}
\end{equation}
The parameter $A$ is equal to the $P$-wave scattering length $a_1^1$ and
the parameter B is related to the slope parameter
$b_1^1=B+4A/(s_\rho-4\mu^2)$. The parameter $s_\rho$ is equal to the $\rho$-mass
squared. Above 970 MeV the $P$-wave amplitude is represented by
the K-matrix parameterization of Hyams {\it et al.}~\cite{hyams73}. The parameters 
$C$ and $D$ of  (\ref{eq:schenk}) are chosen to match both parameterizations
at 969 and \mbox{970 MeV}.
We have checked that the near threshold scalar-isoscalar phase shifts
determined from the differences $\delta_{0}^{0}-\delta_{1}^{1}$ 
in~\cite{pislak01} are insensitive to different $P$-wave
parameterizations. Differences between parameterizations of~\cite{anan}
and~\cite{hyams73} are smaller than the errors 
of $\delta_{0}^{0}-\delta_{1}^{1}$.

The driving terms $d_{\ell}^I(s,s_{max})$ in (\ref{eq:2}) are calculated 
including the contributions of $f_{2}(1270)$ and 
$\rho_{3}(1690)$ and the Regge contributions from the Pomeron, $\rho$-
and \mbox{$f$-exchanges}.
We use the Breit-Wigner parameterization of $f_{2}(1270)$ and 
$\rho_{3}(1690)$ as described in~\cite{kam97}
with masses, widths and $\pi\pi$ branching ratios  taken from~\cite{pdg02}.
The range parameters are chosen to be 5.3~GeV$^{-1}$ and 6.4~GeV$^{-1}$ 
for the $f_2(1270)$ and $\rho_3(1690)$ resonances, respectively 
(see Table~4 of~\cite{hyams73}).
The Regge contributions are parameterized as in~\cite{anan} 
without inclusion of the small $u$-crossed terms.
 The $s_{max}$
limit is set to (2 GeV)$^2$ and our results are fairly close to those
of~\cite{anan}.
The most important contribution in the $\ell=I=0$ channel comes from 
the $f_{2}(1270)$ resonance. It roughly agrees with the result of
Basdevant, Froggatt and Petersen~\cite{basd72} where only the
$f_2(1270)$ contribution was considered and $s_{max}$ was set equal to
(1.46 GeV)$^2$. The $d_{0}^{2}(s,s_{max})$ results are much smaller
than those of~\cite{basd72} due to the difference in $s_{max}$ and to
the lack of the $\rho_{3}(1690)$ contribution in~\cite{basd72}.
 In the $P$-wave there is a
strong cancellation between the contributions of $f_{2}(1270)$ and
$\rho_{3}(1690)$ leading to very small values, again much smaller 
than in~\cite{basd72}. 
We have considered the different parameterizations of $f_2(1270)$ and 
$\rho_3(1690)$ resonances used by~\cite{anan}.
The changes from the Breit-Wigner form, that we used, do not affect the phase shifts 
in all three partial waves by more than one degree below 970 MeV.
In the  $\ell=0,\ I=0$ case the Regge contributions 
are of the order of a 
few percent of the resonance contributions. For the isospin 1 and 2 they are
of the same order as the resonance contributions but the corresponding overall
driving terms are by an order of magnitude smaller than the isospin 0 term.
Alternative Regge contributions were considered following 
papers~\cite{penn73}, \cite{gerard98} and~\cite{zen99}.
These do not change significantly the driving terms, the changes being much 
smaller than one degree in the effective mass range up to 1~GeV.

Our thirteen free parameters, six for the isoscalar $S$-wave, four for the
isotensor one and three for the $P$-wave
 are determined through a least square fit to
the data together with the minimization of the squares of the
differences between the input ({\it ``in''}) and output ({\it ``out''})
 of Roy's equations for the
three waves. We define
\begin{equation}
\chi^2_{total} = \sum_{I=0,1,2} \left[\chi_{exp}^2(I)+\chi^2_{Roy}(I) \right],
\label{eq:chi2total}
\end{equation}
where
\begin{equation}
\chi^2_{exp}(I) = \sum_{i=1}^{N_I} 
\left[ \frac{\sin \left(\delta_{\ell}^I \left(s_i\right) 
- \varphi_{\ell}^I \left(s_i\right) \right)}
       {\Delta \varphi_{\ell}^I \left(s_i\right)} \right]^2.
        \label{eq:chi2exp}
\end{equation}
In  (\ref{eq:chi2exp}) $\varphi_{\ell}^I \left(s_i\right)$ and 
$\Delta \varphi_{\ell}^I \left(s_i\right)$ represent the experimental phase 
shifts and their errors, respectively.
The $\chi^2$ of the fit to Roy's
equations is defined as
\begin{equation}
	\chi^2_{Roy} (I)= \sum_{j=1}^{12} 
	\left [ \frac
   {\mbox{Re }f_{in}^I\left(s_{j}\right)-\mbox{Re }f_{out}^I\left(s_{j}\right)}
	{\Delta f} \right]^2,
        \label{eq:chi2roy}
\end{equation}
where $s_j=[4j+0.001]\mu^2$ for $j=1, ..., 11$ and $s_{12}=46.001\mu^2$.
We take a $\Delta f$ value of $0.5\times 10^{-2}$ to obtain reasonable values of $\chi^2_{Roy}$ 
corresponding to an accuracy of one-half percent. 
Simultaneously we require that the $\chi^2_{exp}$ should not be larger than about 18
for the fit to the 18 data points between 600 and 970~MeV.
We use the CERN MINUIT program which provides errors of the fitted
parameters.

For the isospin 0 wave the number of data points $N_0$ is 24 in  (\ref{eq:chi2exp}). 
It consists of the 18  \df or 18 \uf data points~\cite{kam97} 
between 600 and 970 MeV and of the 
6 points below 400 MeV from the $K_{e4}$ decay experiment~\cite{pislak01}.
In the isotensor wave we use the 12 data points~\cite{hoogland77} from 350 to 1450 MeV.
For the isospin~1 we generate 8 pseudo-data points
using the $K$-matrix parameterization between 600 and 970 MeV~\cite{hyams73}
and we choose $\Delta\varphi^1_1 = 2^{o}$.

\section{Results}

The Pad\'e approximants for the isoscalar $S$-wave amplitude supplemented by 
the model amplitudes~\cite{kamplb97} for $m_{\pi\pi}>970$ MeV together 
with the 
$P$-wave parameterization and that of the isotensor $S$-wave as described above, 
constitute our input to Roy's equations.

A good global fit to data and Roy's equations can be achieved only for the ``down-flat" data. 
The resulting $\ell=0, I=0$ phase shifts, plotted as solid lines in Fig.~1 are
compared to the data.
The $\chi^2$ values are summarized in Table~1. 
For the 18 points between 600 and 960 MeV the $\chi^2$ is 16.6 and it 
is 5.7 for the 6 near threshold points.
The $\chi^2$ per degree of freedom on the isoscalar phase shifts is 1.2.
Roy's equations are very well fulfilled, the differences between the 
real parts {\it ``in"} and {\it ``out"} being smaller 
than $0.8\times 10^{-3}$ for the isoscalar wave and smaller than $2\times 10^{-3}$ for the 
isotensor amplitude.
For the $P$-wave this difference does not exceed $6\times10^{-3}$.
The corresponding parameters are given in Table~2.
The low-energy parameters $a_0^0$, $b_0^0$,  $a^2_0$, $b^2_0$ and $a_1^1$,
compare well within errors with those of~\cite{anan}
but $b_1^1$ is smaller by about $50\%$. 
Relative errors of those parameters vary from about $5\%$ for $a_0^0$, $b_0^0$ and $a_1^1$
to $25\%$ for $b_1^1$.
For the other parameters, they are less than $20\%$ with the exception of $\alpha_2$.
This parameter is negative, so its influence on the values of $\delta_0^0$ is small as can be seen
from  (4).
It furthermore does not appear in the two first coefficients of the low energy expansion of tan $\delta_0^0$.
One then expects the error of $\alpha_2$ to be large.

In the ``up-flat" case a good global fit cannot be obtained, as the 
$\chi^2_{exp}$ on the 18 data points of~\cite{kam97} between 600 and 960 MeV 
is as large as 46.4.
Including the $\chi^2$ of 6.6 from the 6 points of the $K_{e4}$ experiment, 
one obtains the total value equal to 53.0
(see Table~\ref{tab:chi2}).
The $\chi^2$ per degree of freedom on the isoscalar data is as large as 2.9 which shows that 
this fit for the \uf case, plotted as the solid line in Fig.~1b, is not acceptable.  
The agreement with Roy's equations is, however, almost as good as in the \df case.
In Fig.~2 we plot, together with the data, the solid curves of Figs.~1a and
1b representing Roy's fits to the \df and \uf data.
These two curves define a band of isoscalar phase shifts fitting well threshold data and Roy's
equations for two sets of data above 600 MeV.
The ``down"curve reproduces well the \df data while the ``up" curve {\it does not reproduce} the \uf data.

\begin{table}[t!]


\caption[h]{$\chi^2$ values for the different fits; $N_I$ being the number of data and $n_I$ 
that of free parameters, $N_0=24$, $n_0=6$, $N_1=8$, $n_1=3$, $N_2=12$, $n_2=4$.}

\begin{center}
\begin{tabular}{crrr|rrrccrr}
\hline
\hline
~~&  \multicolumn{3}{c|} {\df case} &
\multicolumn{7}{c} {\uf case}  \\
~~&  \multicolumn {3}{c|} {global fit} &
\multicolumn {3}{c} {global fit} &~~& \multicolumn {3}{c} {special fit} \\
\hline
 I           & 
\multicolumn {1}{c}{~~0} & \multicolumn {1}{c}{1}  & \multicolumn {1}{c|}{2} & 
\multicolumn {1}{c}{~~0} & \multicolumn {1}{c}{1}  & \multicolumn {1}{c}{2} &~~& 
\multicolumn {1}{c}{\hspace{0.4cm}0} & \multicolumn {1}{c}{1}  & \multicolumn {1}{c}{2}\\
\hline
 $\chi^2_{exp}(I)$   & 22.3    &   7.0   &   8.1  & 53.0  & 8.4  &  6.9 &~~& \hspace{0.1cm} 19.0 & 7.0 & 8.1    \\
 $\chi^2_{Roy }(I)$  & 0.1   &   6.0   & 0.4 & 0.3   &  7.4  &  0.5 &~~&  
\hspace{1cm} 1.2 $10^4$ & 5.5 & 5.8\\
\hline
 $\chi^2_{total}$ & \multicolumn {3}{c|} { \hst 43.9}  &
                    \multicolumn {3}{c} { 76.5} &~~&
                    \multicolumn {3}{c} { \hspace{2.45cm} 1.2 $10^4$}\\
\hline
\hline 
\end{tabular}
\end{center}
\label{tab:chi2}
\end{table}


\begin{table}[t!]

\caption[h]{Parameters from the fit to Roy's equations and to the \df data of~\cite{kam97};
$\alpha_3$, $\alpha_7$, $\alpha_9$, B, C, D, $\Lambda_1$ and $\Lambda_2$ are dependent parameters.}


\begin{center}
   \begin{tabular}{cll|cll|cll}
\hline
\hline
\multicolumn {3}{c|} {isoscalar $S$-wave} &  
\multicolumn {3}{c} {isovector $P$-wave} &  
\multicolumn {3}{|c} {isotensor $S$-wave} \\ 
\hline 
\hsl       $a_{0}^0$ & ~$0.224\pm 0.013$ &   &
       $a^1_1$ &  \hst  ~$\,(3.96  \pm 0.24)\,10^{-2}$ &  \hsmu $\mu^{-2} $ &  
        $a_{0}^2$ & \hst  $(-3.43\pm 0.36)\,\, 10^{-2}$  &  \\
\hsl       $b_{0}^0$ & ~$0.252\,\,^{+\,0.012}_{-\,0.010}$ &  \hsmu $\mu^{-2}$  &
        $b^1_{1}$ & \hst  ~$\,(2.63\,\,^{+\,0.67}_{-\,0.66})\,\,10^{-3}$ & \hsmu $\mu^{-4}$  &
        $b_{0}^2$ & \hst  $(-7.49\,\,^{+\,1.01}_{-\,1.65})\,\,10^{-2}$ &  \hsmu \hst $\mu^{-2}$  \\
\hsl       $\alpha_{2}$ & \hst $-4.41\,\,\,\,\,^{+\,2.06}_{-\,5.05}$ &  \hsmu $\mu^2$  &
       $s_\rho$ & \hst ~$30.87   \pm 0.14$ &  \hsmu $\mu^2\,\,\,$ &
        $\beta_{1}$ & ~~$4.88\pm 0.25$ & \hsmu \hst $\mu\,\,\,\,\,\,$  \\
\hsl       $\alpha_{4}$ & ~$7.53\,\,\,\pm 0.32$ & \hsmu $\mu^2$  &
         &   &   &
	$\beta_{2}$ & ~~$1.23\pm 0.21$ & \hsmu \hst $\mu\,\,\,\,\,\,$  \\
\hsl       $\alpha_{5}$ & \hst ~$\,(2.29\;\;\,^{+\,0.35}_{-\,0.45})\;10^{-2}$ &  \hsmu $\mu^{-5}$  &
         &   &   &
         &   &    \\
\hsl       $\alpha_{6}$ & \hst ~$12.0\;\,\,\,\,\,\,^{+\,0.5}_{-\,0.3}$ &  \hsmu $\mu^2\,\,\,$   &
         &   &   &
         &   &    \\
\hline
\hsl       $\alpha_{3}$ & \hst $-0.153$ &  \hsmu $\mu^{-3}$  &
       $B $  &    \hst   $ -3.27\,\, 10^{-3}$ & \hsmu $\mu^{-4} $   &
        $\Lambda_{1}$ & ~~$0.268$ & \\
\hsl       $\alpha_{7}$ & \hst $-0.320\,\,10^{-3}$ &  \hsmu $\mu^{-7}$  &
        $C$ & ~$5.24\,\,10^{-4} $ &   \hsmu $\mu^{-6}$  &
	$\Lambda_{2}$ & \hst  ~$-0.0257$ & \\
\hsl       $\alpha_{9}$ & \hst $-0.312\,\,10^{-4}$ & \hsmu $\mu^{-9}$  &
        $D$ & \hst  $-2.66\,\,  10^{-5}$ &  \hsmu $\mu^{-8}$  &
         &   &    \\
\hline
\hline
   \end{tabular} 
\end{center}    
\label{allparameters}
\end{table}

\begin{figure}[h!]
\label{padebands}
\mbox{\epsfxsize 14cm\epsfysize 19cm\epsfbox{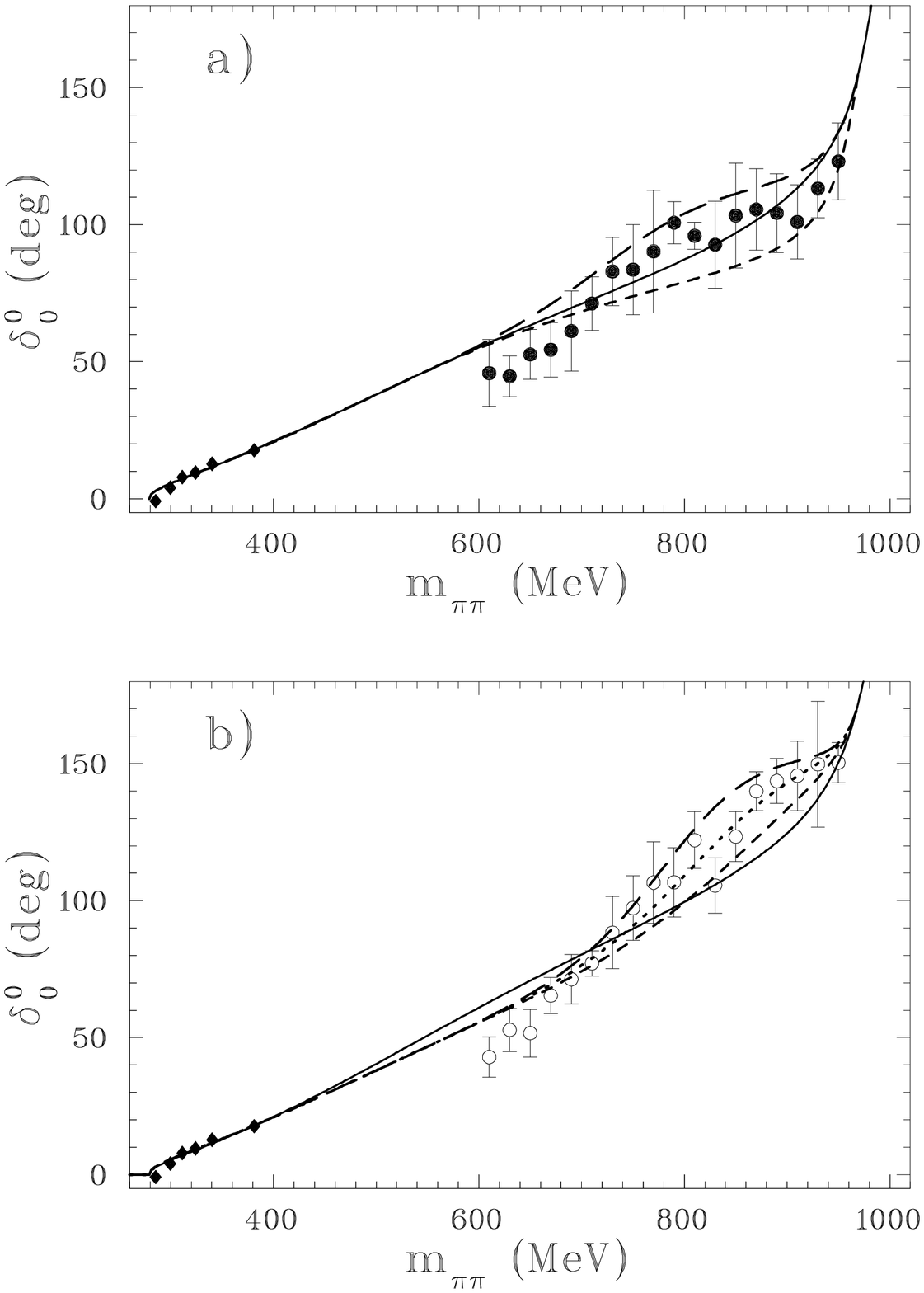}}\\
{Fig. 1: Scalar-isoscalar phase shifts for \textbf{a)} \df data (full circles) and 
\textbf{b)} \uf data (open circles)~\cite{kam97}. 
Diamonds denote the K$_{e4}$ data~\cite{pislak01}.
Solid lines represent fits to Roy's equations and to data.
Dashed lines represent bands of the Pad\'e fits to the data of~\cite{kam97}.
The dotted line in \textbf{b)} shows the special Pad\'e fit.}
\end{figure}   

\begin{figure}[h!]
\label{oneband}
\mbox{\epsfxsize 14cm\epsfysize 9.cm\epsfbox{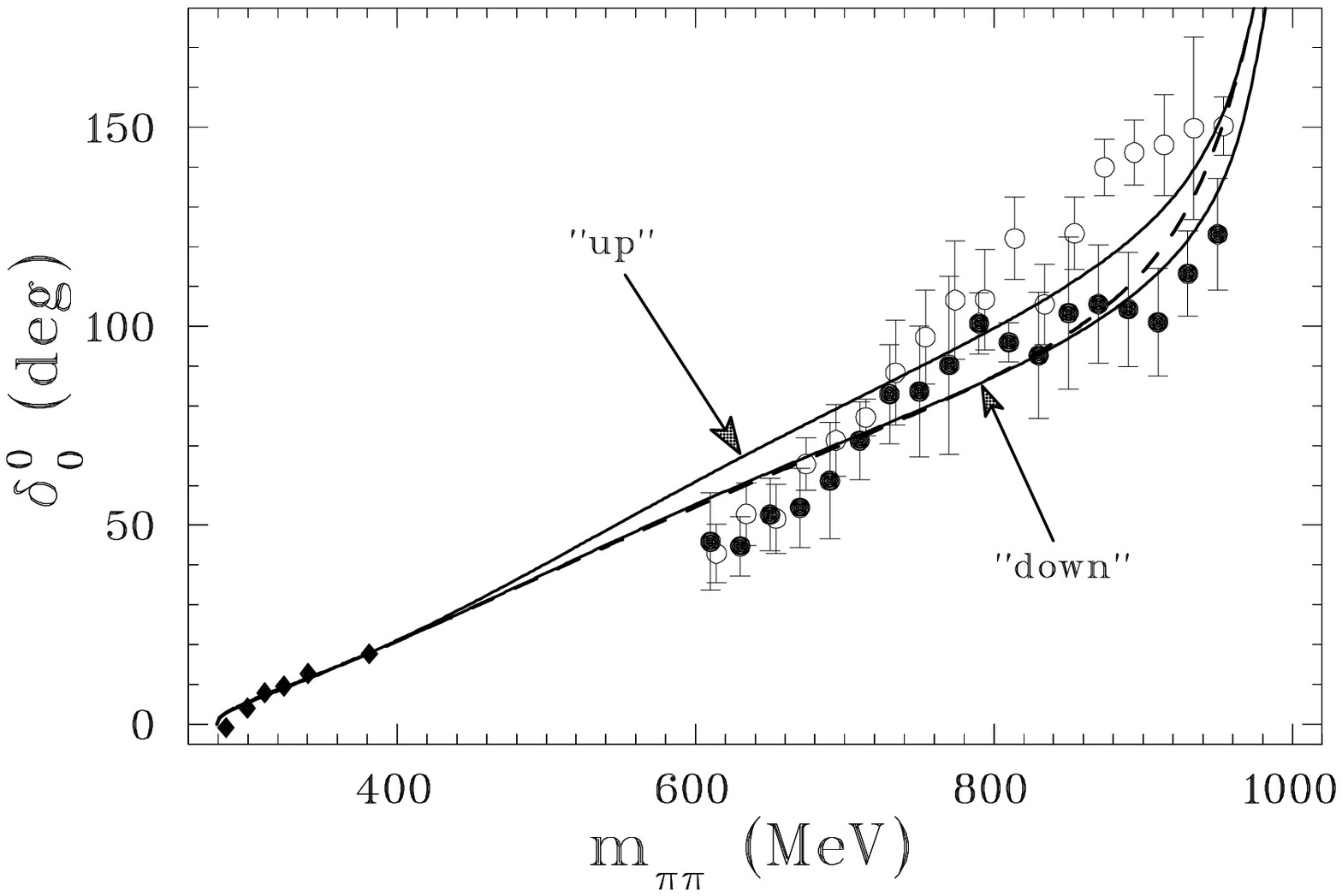}}\\
{Fig. 2: Solid lines -- fits to Roy's equations and to the \df data (full circles) and to the \uf
data (open circles) of~\cite{kam97} as well as to  the $K_{e4}$ data of~\cite{pislak01}
represented by diamonds.
The dashed curve is Roy's fit to the $K_{e4}$ data and to the \uf data in the restricted range of
\mpp up to 740 MeV.}
\end{figure}

The main differences between the ``down-flat" and ``up-flat" solutions 
are for energies between 800 and 970 MeV. 
Below 740 MeV phase shifts of both solutions are compatible within their error bars
and the ``down" curve reproduces well the \uf data (see Fig.~2).
We have checked that there is a possibility to find a good fit to Roy's equations and to the \uf data
in this limited range of effective mass.
This is shown as the dashed line in Fig.~2.
Differences of phase shifts between this line and the ``down" one are smaller than one degree up to
840 MeV.
Therefore, based on the \df fit, we build up a special Pad\'e fit to the
``up-flat" data in order to reproduce as well as possible their $m_{\pi\pi}$ dependence.
We proceed in the following way.
Below 600 MeV this fit is constrained to approximate the previously 
obtained \df isoscalar amplitude and in particular to reproduce its 
scattering length $a^0_0$ and slope parameter $b^0_0$ as well as its two values at 
500 and 550 MeV.
The corresponding $\chi^2_{exp}(0)$ of 19.0 is good but the 
$\chi^2_{Roy}(0)$ of $1.2\times 10^4$ is very large.
The differences between the 
{\it ``in"} and {\it ``out"} real parts are as large as 0.25 around 900 MeV.
So, if one tries to improve the fit to the ``up-flat" data then 
one spoils the fit to Roy's equations.

\begin{figure}[htbp]
\label{banddown}
\mbox{\epsfxsize 13cm\epsfysize 19cm\epsfbox{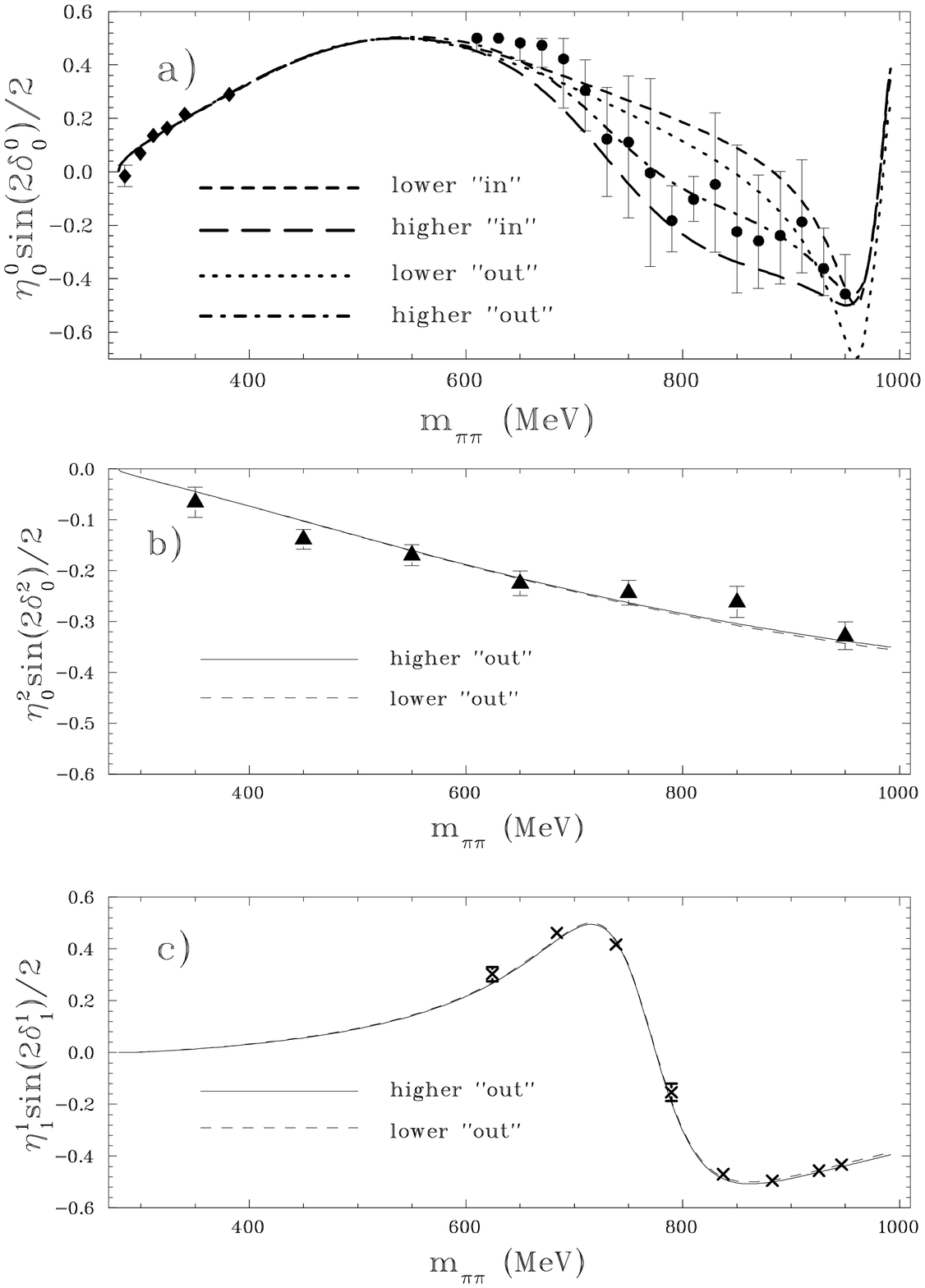}}\\   
{Fig.~3: Real parts of the \pipi amplitudes (multiplied by $2ks^{-1/2}$)
corresponding to the \df data~\cite{kam97} (full circles).
In {\bf a)} input and output bands of the scalar-isoscalar real parts are shown. 
Diamonds denote the $K_{e4}$ data~\cite{pislak01}.
Solid and dashed lines in {\bf b)} and {\bf c)} represent the output bands.
In {\bf b)} triangles denote the isotensor data of~\cite{hoogland77}.
Crosses in {\bf c)} are the isovector pseudo-data calculated from the $K$-matrix fit of~\cite{hyams73}.
}    
\end{figure}   

\begin{figure}[htbp]\centering   
\label{bandup}
\mbox{\epsfxsize 13cm\epsfysize 19cm\epsfbox{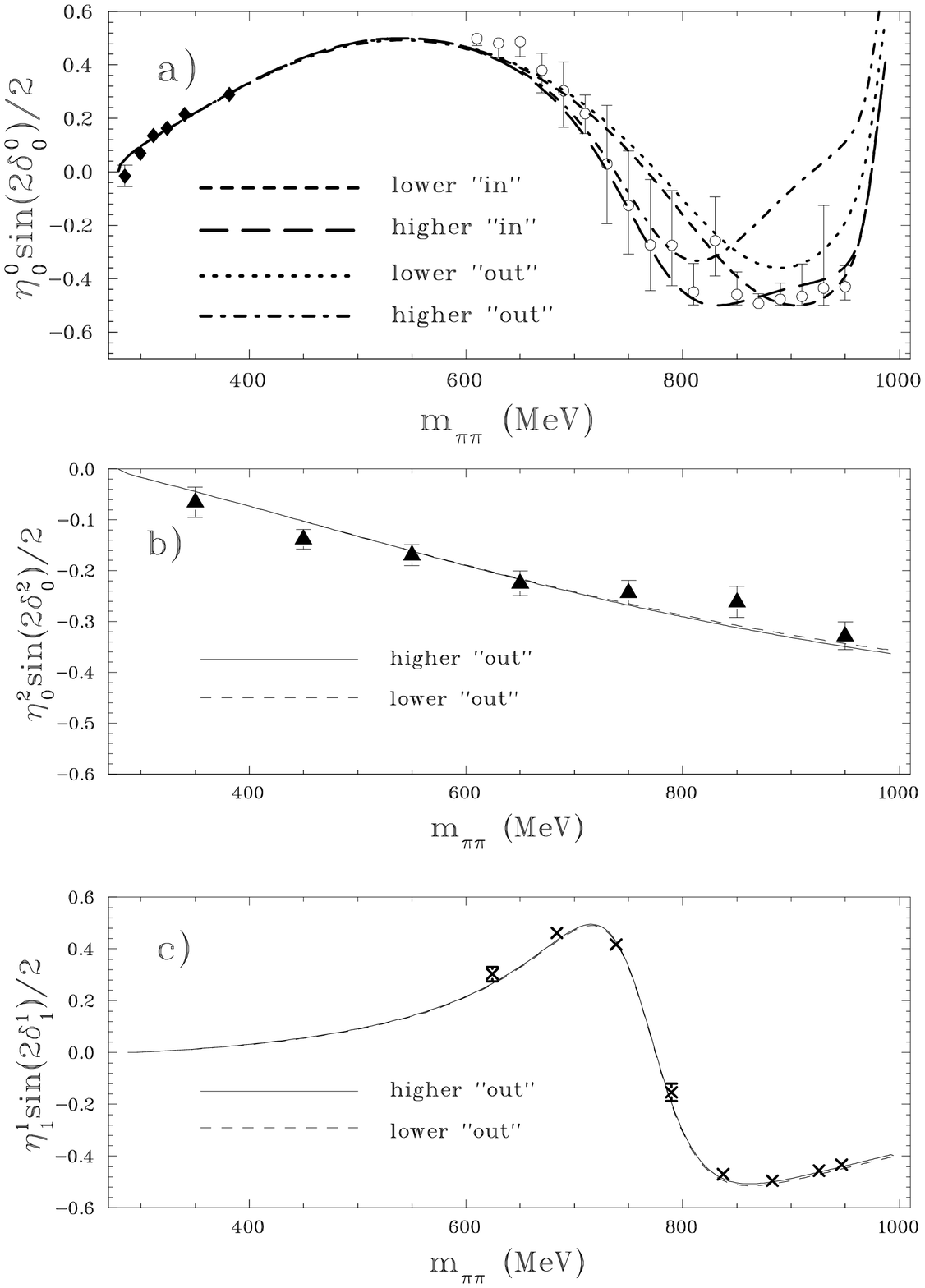}}\\   
{Fig.~4: Same as in Fig.~3, but for the ``up-flat" data 
of~\cite{kam97}}    
\end{figure}   

We have studied the influence of the experimental errors on the 
$\pi\pi$ input amplitudes by calculating Roy's equations for two 
extreme isoscalar amplitudes fitted to the data points shifted 
upwards (``higher-in") or downwards (``lower-in") by their errors.
Below 600 MeV these fits were constrained in the same way as in 
the special Pad\'e fit to the ``up-flat" data just described above.
The corresponding long- and short-dashed lines in Figs.~1a and~1b show typical 
bands delimiting the possible values of the 
experimental data within their errors.
The resulting output curves, lower ``out" and higher ``out", of the numerical integrations 
of Roy's equations 
for the three partial-wave real parts, multiplied by $2ks^{-1/2}$,
are displayed in Figs.~3 and~4.
In Figs.~3a and 4a we also show the input band for the real part of the isoscalar wave, 
limited by two dashed lines, called lower ``in" and higher ``in". 
These dashed lines correspond to the dashed lines shown in Figs.~1a and 1b.
We do not show the {\it ``in"} lines in Figs.~3b,~3c, 4b and~4c since they are 
almost~indistinguishable from the {\it ``out"} lines.

Below 600~MeV  Roy's equations are 
well  satisfied for all waves.
At higher energies a good agreement is found 
for the $P$-wave and $I=2$ $S$-wave. 
As seen in Fig.~3, in the ``down-flat" case both {\it ``out"} curves lie inside the 
band limited by the {\it ``in"} curves up to 937 MeV. 
Above this energy the lower ``out" curve starts to lie below the higher ``in" and at about 
950 MeV the higher ``out" starts to be above the lower ``in" curve.
The {\it ``in"} band is then inside the {\it ``out"}  one.
It means that there is still a possibility to find an {\it ``out"} solution within the 
{\it ``in"} band.
We can then conclude that within the error bars the \df solution satisfies Roy's
equations and consequently crossing symmetry. 
On the contrary, in Fig.~4, for the scalar-isoscalar \uf  solution, 
the output band lies
outside the input one from~840 to~970 MeV. This eliminates
the \uf solution as it does not satisfy crossing symmetry in
that energy range.

\section{ Conclusions }

 We have studied the scalar-isoscalar $\pi\pi$ 
amplitudes constructed from two sets of phenomenological phase shifts 
called ``up-flat" and ``down-flat"~\cite{kam97}.
It has been shown that only the ``down-flat" solution satisfies well crossing symmetry. 
The conclusion about violation 
of Roy's equations for the ``up-flat" 
data is insensitive to the variation of the different parameters in our 
input: reasonable changes of the $\ell=1,\ I=1$ and $\ell=0,\ I=2$ amplitudes,
shift of the upper integration limit $s_{max}$ from (2 GeV)$^2$ to 
(1.46 GeV)$^2$, modifications of the $f_2(1270)$ and $\rho_3(1690)$ resonance parameters and 
of the Regge amplitudes in the driving terms. All these changes do not
lead to complete overlap of the 
input and output bands for the ``up-flat" solution.

Recently a joint analysis of the $S$-wave $\pi^+\pi^-$ 
data~\cite{grayer75,becker79} and of the new $\pi^{0}\pi^{0}$ data~\cite{gunter01}, 
obtained by the E852 Collaboration at 18.3 GeV/c, has been performed~\cite{kam01}.
It has been shown that, using the one-pion and $a_{1}$-exchange model 
developed in~\cite{kam97}, the calculated $S$-wave intensity of the 
$\pi^{0}\pi^{0}$ system agrees with the measured $\pi^{0}\pi^{0}$ 
intensity only for the $\pi\pi$ amplitudes obtained from the 
``down-flat" phase shifts.
 So, in~\cite{kam01}   the ``up-flat" solution has 
 also  been eliminated and a new 
``down-flat" set of phase shifts, compatible with the $\pi^+\pi^-$ and 
$\pi^{0}\pi^{0}$ data, has been found.
We have verified that these new ``down-flat" phase shifts, 
parameterized as described above, satisfy well 
Roy's equations.

To conclude, using the theoretical constraints of crossing symmetry, 
unitarity and analyticity, the four-fold ambiguities in the 
isoscalar $S$-wave $\pi\pi$ phase shifts, in the  $\pi\pi$
invariant  mass range from 800  to 1000 MeV, have been eliminated 
in favour of the \df solution.

\vspace{.5cm}

\textbf{Acknowledgments}

We acknowledge very useful correspondence on the calculation of the
driving terms with H.~Leutwyler and B.~Ananthanarayan. We thank
B.~Nicolescu and P.~\.Zenczykowski for fruitful discussions on Regge
amplitudes. We are grateful to K.~Rybicki for helpful collaboration on
the $\pi \pi$ phase shifts and to G.~Colangelo,  J.~Gasser,  B.~Moussallam and
J.~Stern for enlightening comments. This work has been performed in the
framework of the IN2P3-Polish Laboratories Convention (project number 99-97).

\end{document}